\documentclass[final,5p,times,twocolumn]{elsarticle}

\usepackage{amssymb}
\usepackage{amsmath}
\usepackage{amsfonts}
\usepackage{graphicx}
\usepackage{textcomp}
\usepackage{algorithm}
\usepackage{algpseudocode}
\usepackage{array}
\usepackage{stfloats}
\usepackage{url}
\usepackage{verbatim}
\usepackage{hyperref}
\usepackage{booktabs}
\usepackage{subcaption}

\usepackage{natbib}
\journal{International Journal of Electrical Power & Energy Systems}

\begin{document}

\begin{frontmatter}



\title{Resilience of IEC 61850 Sampled Values-Based Protection Systems Under Coordinated False Data Injections}



\author[label1]{Denys Mishchenko} 

\affiliation[label1]{organization={The Department for Electric Energy, Norwegian University of Science and Technology},
            addressline={O.S. Bragstads plass 2a}, 
            city={Trondheim},
            postcode={NO-7491},
            country={Norway}}

\author[label1]{Irina Oleinikova} 

\author[label2]{L{\'a}szl{\'o} Erd{\H o}di} 
\affiliation[label2]{organization={The Department of Information Security and Communication Technology, Norwegian University of Science and Technology},
            addressline={O.S. Bragstads plass 2a}, 
            city={Trondheim},
            postcode={NO-7491},
            country={Norway}}

\begin{abstract}
This paper assesses the resilience of IEC 61850 digital substations under False Data Injection Attacks (FDIAs) targeting the Sampled Values (SV) protocol. The multicast nature of SV, while enabling time-critical automation, exposes substations to cyber intrusions capable of disrupting protection functions and causing large-scale outages. To evaluate these risks, coordinated attack vectors involving both physical and cyber access at the bay level are experimentally analyzed using an advanced setup based on industrial-grade intelligent electronic devices (IEDs). The proposed attacks simultaneously manipulate multiple electrical parameters in a coordinated and physically consistent manner.

Experimental results confirm the feasibility of stealthy multi-vector FDIAs that can trigger false protection actions, conceal real faults, or block protection mechanisms while maintaining realistic signal behavior. The Power Hardware-in-the-Loop (PHIL) testbed enables closed-loop evaluation under strict timing, communication, and protection logic constraints, reflecting real device behavior beyond simulation and controller-level HIL environments. The findings reveal critical vulnerabilities in SV-based protection schemes that directly affect grid reliability, particularly under realistic attacker positioning.

To address these challenges, a defense strategy covering deterrence, prevention, detection, mitigation, and resilience is analyzed, with emphasis on bay-level infrastructure. Furthermore, a resilience-oriented method based on trusted independent channels and cross-verification of SV data within the protection logic is outlined as a complementary countermeasure for scenarios where existing standardized security mechanisms are insufficient.
\end{abstract}



\begin{keyword}
smart grid, IEC 61850, sampled values, false data injection attack, cybersecurity.
\end{keyword}

\end{frontmatter}




\section{Introduction}
\subsection{Background}

Modern power systems rely on digital communication and automation to enable fast and coordinated protection functions \cite{nazari2023impact}. The IEC~61850 standard supports this paradigm by defining interoperable communication among Intelligent Electronic Devices (IEDs), including protocols such as Sampled Values (SV) for time-critical measurements and Generic Object-Oriented Substation Events (GOOSE) for event-based signaling \cite{IEC61850General}. While these protocols enhance performance and flexibility, their multicast nature and limited native security mechanisms introduce potential cyber-physical attack surfaces.

At the bay level, protection relays depend on accurate current and voltage measurements as well as precise time synchronization for correct operation. Consequently, manipulation of SV traffic or disruption of synchronization services can lead to false trips, fault masking, or blocking of protection functions. An adversary with access to the substation communication network—either through compromised devices or unauthorized connections—can exploit these vulnerabilities and impact protection behavior, potentially leading to cascading disturbances \cite{makrakis2021industrial, el2019iec}.

This work investigates these vulnerabilities through scenario-based analysis of advanced False Data Injection Attacks (FDIAs) targeting SV communication and related services, providing experimental insight into their impact on protection systems and motivating resilience-oriented approaches.

\subsection{Related work}

Existing research on cybersecurity in IEC 61850-based substations has been extensively studied in recent years. Analysis of the literature reveals several recurring gaps, which are summarized in Table~\ref{table:1} and discussed in detail below. These gaps can be viewed from four perspectives: (1) most studies primarily focus on detection and protection mechanisms, while attack analysis is treated as a secondary validation scenario; (2) the Power Hardware-in-the-Loop (PHIL)-based experimental setups with industrial-grade IEDs, representing closed-loop interaction between simulated systems and physical protection devices, are rarely considered; (3) many contributions address IEC 61850 security at a general level, with limited focus on SV-specific communication; (4) false data injection attacks are typically modeled as single-parameter perturbations, whereas coordinated multi-parameter manipulation preserving physical consistency is seldom investigated.

Addressing  on detection and protection mechanisms Presekal et al. \cite{presekal2023attack} proposed a hybrid deep learning approach for anomaly detection in operational technology networks, while Ustun et al. \cite{ustun2021artificial} developed a machine learning-based intrusion detection system for identifying FDIAs under fault conditions. Esiner et al. \cite{esiner2022lomos} introduced lightweight cryptographic signatures for securing GOOSE and SV communication. Despite these contributions, attack models in such studies are typically simplified and primarily used for validation, rather than being systematically analyzed in terms of feasibility and impact on protection systems.

\begin{table*}[!htbp]
\centering
\caption{\textbf{Summary of related works on IEC 61850 cybersecurity}}
\label{table:1}
\scalebox{0.8}{
\begin{tabular}{|p{105pt}|p{100pt}|p{70pt}|p{50pt}|p{50pt}|p{150pt}|}
\hline
\textbf{Author} & \textbf{Experimental Setup} & \textbf{Protocol(s)} & \textbf{Multi-vector attacks}  & \textbf{Physically Consistent Attacks} & \textbf{Proposed countermeasure} \\
\hline
Mocanu et al. \cite{mocanu2021real} & HIL (industrial IEDs) & SV & $\times$ & $\times$ & IDS  \\
\hline
Hong et al. \cite{hong2021implementation} & HIL (virtual IEDs) & SV & $\times$ & Partial & MAC auth. \\
\hline
Hussain et al. \cite{hussain2023effective} & HIL (virtual IEDs) & SV & $\times$ & $\times$ & MAC auth. \\
\hline
Rodriguez et al. \cite{rodriguez2021fixed} & HIL (virtual IEDs) & GOOSE, SV & $\times$ & $\times$ & AES-GCM (FPGA) \\
\hline
Elrawi et al. \cite{elrawy2024geometrical} & Simulation & GOOSE & $\times$ & $\times$ & Geometric auth./enc \\
\hline
Ustun et al. \cite{ustun2021artificial} & Simulation & SV & $\times$ & $\times$ & ML-based IDS  \\
\hline
Esiner et al. \cite{esiner2022lomos} & Simulation & GOOSE, SV & $\times$ & $\times$ & LoMoS signatures \\
\hline
Rajkumar et al. \cite{rajkumar2024dynamical} & Simulation & GOOSE, SV & Partial & $\times$ & Cascading failure analysis \\
\hline
Farooq et al. \cite{farooq2019s} & Simulation & GOOSE, SV & $\times$ & $\times$ & S-GoSV framework \\
\hline
Presekal et al. \cite{presekal2023attack} & Simulation & OT protocols & $\times$ & $\times$ & Hybrid DL IDS \\
\hline
\textbf{This study} & \textbf{PHIL (industrial IEDs)} & \textbf{SV, PTP} & $\checkmark$ & $\checkmark$ & \textbf{Resilience-oriented extension} \\
\hline
\end{tabular}
}
\vspace{-15pt}
\end{table*}

Another important aspect concerns experimental validation. Many studies rely on simulation or Hardware-in-the-Loop (HIL) environments with abstracted or virtualized devices. Hong et al. \cite{hong2021implementation} implemented IEC 62351-6 authentication schemes in an HIL testbed, while Hussain et al. \cite{hussain2023effective} proposed a MAC-based security mechanism validated on hardware platforms. Mocanu et al. \cite{mocanu2021real} investigated SV communication performance under cyberattacks using HIL-based setups. However, the majority of these works do not fully capture the timing constraints, communication behavior, and protection logic associated with industrial-grade IEDs, limiting the assessment of attack feasibility under realistic conditions.

In terms of communication protocols, many contributions address IEC 61850 security at a general level or focus primarily on GOOSE messaging. Rodriguez et al. \cite{rodriguez2021fixed} developed a low-latency FPGA-based cryptographic solution for GOOSE and SV protection, while Farooq et al. \cite{farooq2019s} proposed a framework for secure message generation using digital signatures. Nevertheless, detailed analysis of SV-specific attack execution, including its timing and multicast characteristics, remains limited.

Finally, the majority of existing works consider simplified false data injection attacks. Rajkumar et al. \cite{rajkumar2024dynamical} analyzed cascading failures caused by cyberattacks in power systems, and Elrawi et al. \cite{elrawy2024geometrical} proposed a geometrical approach for secure communication. In these and similar studies, attacks typically involve single-parameter manipulation or minor perturbations designed to trigger protection thresholds. Such assumptions do not reflect realistic grid behavior, where electrical parameters are interdependent. As a result, detection and protection mechanisms validated under these conditions may not generalize to coordinated attacks that preserve physical consistency.

Existing studies predominantly emphasize protection and detection strategies, often relying on simplified attack models and non-realistic validation environments. In particular, limited attention is given to coordinated multi-vector attacks, physically consistent manipulation of measurements, and experimental validation under real industrial constraints. These gaps restrict a comprehensive assessment of actual vulnerabilities in IEC~61850 SV-based protection systems.

\subsection{Contribution}

This paper presents an evaluation of coordinated FDIAs targeting the IEC~61850 SV protocol, with focus on protection system behavior under realistic operating conditions. The study is conducted using industrial-grade setup enabling analysis of attack feasibility under practical constraints.

The main contributions of this work are summarized as follows:
\begin{itemize}
    \item \textbf{PHIL-based experimental validation}: Demonstration of FDIAs on SV communication using a closed-loop PHIL testbed with industrial-grade protection IEDs and a physical communication network, capturing timing constraints, protocol behavior, and protection logic of deployed substation systems.
    
    \item \textbf{Coordinated multi-parameter attack modeling}: Development and experimental validation of coordinated attack strategies that simultaneously manipulate multiple electrical parameters while preserving physical consistency, enabling the emulation of realistic fault conditions.
    
    \item \textbf{Feasibility and impact assessment}: Experimental analysis of attack execution and its impact on protection system behavior, including false tripping, fault masking, and blocking of protection functions under stealthy conditions.
    
    \item \textbf{Evaluation of defense limitations}: Targeted assessment of existing protection and cybersecurity mechanisms under the implemented scenarios, highlighting their limitations in detecting coordinated and physically consistent attacks.
    
    \item \textbf{Resilience-oriented enhancement}: Proposal of a complementary protection concept based on cross-verification using an independent trusted signal, designed to improve resilience against advanced FDIAs.
\end{itemize}

\subsection{Paper Structure}

The paper is structured as follows. Section~\ref{sec:system_model} describes the system model, including the physical and communication layers of an IEC~61850-based digital substation. Section~\ref{sec:attack_scenarios} presents the considered coordinated false data injection attack scenarios targeting the SV protocol. Section~\ref{sec:results_analysis} details the experimental setup based on a hardware-in-the-loop testbed with industrial IEDs and analyzes the impact of the implemented attacks. Section~\ref{sec:proposed_measures} discusses the evaluated defense mechanisms and introduces the resilience-oriented approach. Finally, Section~\ref{sec:conclusion} concludes the paper and outlines future research directions.

\section{System Model}\label{sec:system_model}

This section introduces the architecture of an IEC 61850-based digital substation, structured into physical and cyber layers. The physical layer covers the hierarchical levels of the substation and associated network infrastructure, while the cyber layer defines protocols and data flows for automation and protection. Both layers are discussed with emphasis on their role in security exposure.

\subsection{Physical layer}

\begin{figure}[!htbp]
\centering
\includegraphics[width=1\columnwidth]{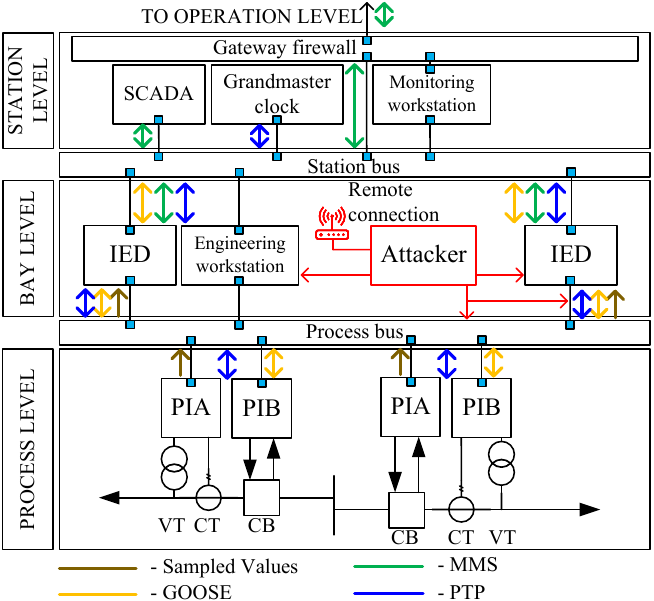}
\caption{Cyber--physical architecture of an IEC~61850-based digital substation}
\label{fig:Structure}
\end{figure}

The physical layer comprises three hierarchical levels—process, bay, and station—interconnected via the process and station buses, as illustrated in Fig.~\ref{fig:Structure}. At the process level, process interface analog (PIA) devices, such as merging units (MUs), digitize analog measurements from current transformers (CTs) and voltage transformers (VTs) into IEC~61850-9-2–compliant SV \cite{IEC61850-9-2}. In parallel, process interface binary (PIB) devices support fast circuit breaker (CB) actuation. At the bay level, protection relays process SV streams for fault detection and breaker control, while engineering workstations provide configuration and supervision. At the station level, supervisory systems, including SCADA and the Grandmaster (GM) clock, aggregate bay-level information, interface with control centers, and distribute precise time synchronization via PTP.

From a networking perspective, process and station buses may adopt star, ring, or double-ring topologies \cite{mocanu2021real}. While star configurations offer simplicity, they introduce single points of failure, whereas ring and double-ring architectures improve redundancy at the expense of increased complexity and cost. These design trade-offs influence both operational reliability and the available attack surface.

The described components and configurations may introduce vulnerabilities that elevate cybersecurity risk in substation environments. Engineering workstations connected to external networks can serve as entry points for malware or unauthorized access. Unpatched device firmware, insecure switch configurations, and insufficient network segmentation (e.g., absence of demilitarized zones between OT and IT domains) further increase the risk of lateral movement. In addition, compromised supply-chain components or inadequate physical protection—such as unauthorized access to network switches—can expose critical assets \cite{cisa2023,cisa2024}. Since process-level devices provide time-critical measurements and PTP underpins synchronized protection operation, compromises at this layer can directly undermine protection integrity and grid stability.

\subsection{Cyber layer}

The cyber layer comprises the time-critical process bus (bay--process) and the station bus (bay--station), and it determines the data formats, timing constraints, and attack surface on which protection logic depends. The process bus carries IEC~61850 SV as Layer-2 multicast frames delivering digitized phasors, instantaneous currents and voltages, and associated meta-fields (e.g., smpCnt, the SV sample counter, and frame sequence), along with GOOSE messages for fast peer-to-peer signaling. As SV and GOOSE typically transit unencrypted at Layer~2 and rely on multicast delivery, an adversary with local network access can readily sniff, inject, or replay frames. Manipulation of SV frame continuity directly affects relay measurement integrity and protection trip decisions \cite{maurya2024analysis}.

The station bus supports MMS over TCP/IP for information exchange between IEDs and higher-level devices. MMS enables access to configuration data and SCL content which, if exposed or modified, may facilitate higher-privilege attacks, such as altering protection settings or logic parameters. Consequently, weak station-level authentication or compromised engineering workstations can significantly expand an attacker’s capabilities beyond basic frame manipulation.

Process and station buses are commonly segmented using VLANs to reduce congestion and logically isolate traffic. However, VLANs represent a logical rather than physical separation. Misconfiguration, VLAN hopping, or switch compromise may expose data streams to adversaries even when nominally isolated \cite{sundararajan2018survey}. In combination with the multicast nature of SV and GOOSE, such weaknesses can enlarge the effective attack surface.

Time synchronization is provided by the GM via PTP and is essential for consistent event ordering across protection devices. Attacks on PTP, including GM spoofing, offset injection, selective delay, or DoS, can introduce timestamp drift and smpCnt inconsistencies, leading to disagreement among IEDs and potential blocking or miscoordination of protection functions \cite{akbarzadeh2023attacking}. Practical countermeasures include PTP redundancy, sanity checks on timestamp variations, and authenticated PTP where supported; however, these measures increase deployment complexity and must be evaluated against the strict timing constraints of SV-based protection.

Such vulnerabilities may enable unauthorized access to otherwise isolated traffic streams, broadening the attacker’s reach. When exploited through coordinated attack strategies, these expanded vectors can amplify the impact of intrusions initiated even at the bay level, motivating the need for resilience-oriented protection analysis.

\section{Attack Scenarios}\label{sec:attack_scenarios}

\begin{figure}[!htbp]
\centering
\includegraphics[width=1\columnwidth]{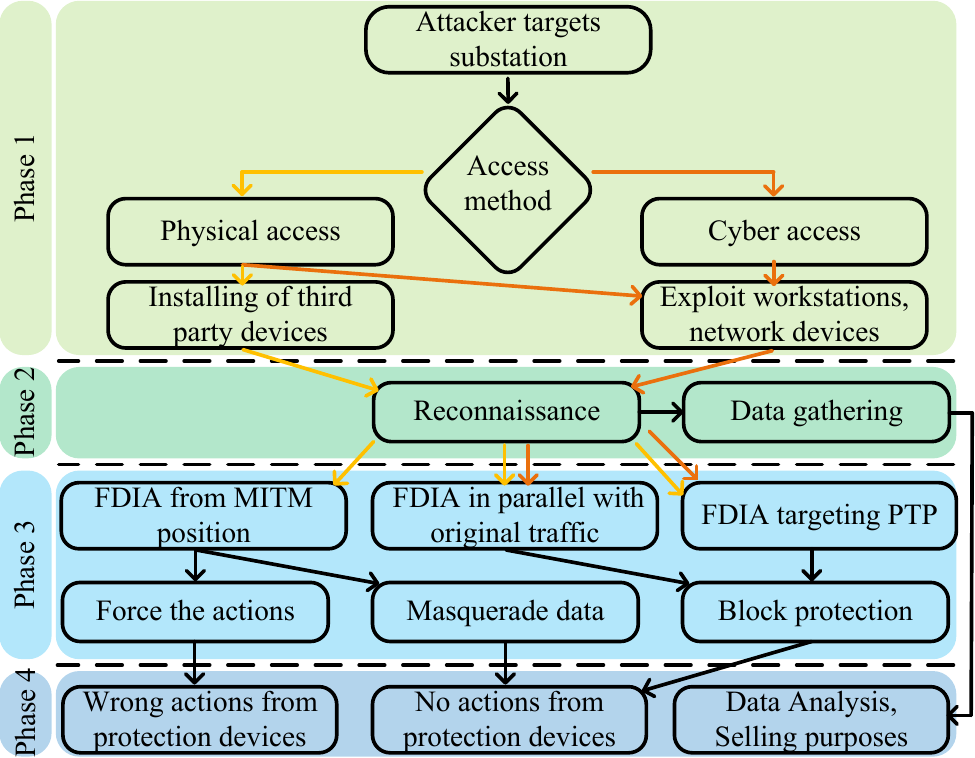}
\caption{Full attacker lifecycle showing possible actions and expected consequences at each stage}
\label{fig:flowchart}
\end{figure}

This section outlines attack scenarios targeting an IEC~61850-based substation automation system, with emphasis on advanced FDIAs affecting the SV protocol. The scenarios are structured into four phases from the attacker’s perspective: initaial access, preparation, FDIA execution, and attacks impact. Each phase comprises specific actions that exploit vulnerabilities across the physical and cyber layers, as illustrated in Fig.~\ref{fig:flowchart}. 

\subsection{Phase 1 — Initial access}

This phase begins when an adversary establishes an initial foothold in the substation network through either physical insertion or remote compromise. In a physical insertion scenario, the adversary may exploit weak site protection, such as insufficient surveillance, inadequate access control, or unsecured maintenance ports, to connect a third-party device (e.g., a single-board computer) to the local network. The device may be attached to an unused switch port or placed inline between communicating units, as shown in Fig.~\ref{fig:Structure}. From this position, the adversary can observe traffic, enable port mirroring, manipulate VLAN tags, clone legitimate MAC addresses, or selectively forward and modify frames so that injected traffic blends with normal communication. To remain stealthy, the adversary may preserve critical header fields and timing characteristics, such as expected smpCnt progression and inter-frame spacing for SV frames, while avoiding link errors or anomalous control-plane activity.

Alternatively, a remote compromise path may originate from an engineering workstation with weak authentication or unpatched software. Once compromised, such a workstation may be used to reconfigure switches, enable port mirroring, or inject forged traffic through otherwise legitimate interfaces. Although VLANs are intended to isolate traffic, misconfigurations or improper enforcement may allow an adversary with a single foothold to gain visibility into additional segments. Together, these physical and remote entry paths can provide the visibility and limited control required to proceed to reconnaissance and the execution of FDIAs.

\subsection{Phase 2 — Preparation}

After establishing a foothold, the adversary may conduct reconnaissance to understand the substation topology, device roles, and timing behavior prior to launching attacks. This phase typically combines two complementary strategies: passive traffic observation and limited active probing \cite{mishch}. Passive reconnaissance focuses on intercepting data frames to identify publishers, multicast groups, sampling rates, smpCnt progression, and typical measurement ranges. Such information enables the construction of protocol-compliant forgeries with low detectability. Since SV and GOOSE provide limited contextual information, the adversary may additionally monitor other unencrypted sensor or actuator channels to infer node configuration and physical location \cite{PowerTech2025}.

Active reconnaissance involves carefully paced queries or benign management operations to retrieve device configurations, network settings, and the functional role of individual nodes within the substation. While active probing increases the risk of detection, it yields richer configuration metadata that directly informs the selection of FDIA parameters and stealth constraints applied during the execution phase.

\subsection{Phase 3 — Execution of FDIAs}

During the execution phase, the adversary—positioned at the bay level—targets time-critical communication protocols such as SV and PTP. A successful FDIA on SV manipulates measurement data exchanged between IEDs, potentially triggering incorrect disconnections within the substation. If falsified measurements are accepted as legitimate, they may further propagate through protection schemes \cite{hou2008iec} or reach higher-level systems via MMS, resulting in erroneous control actions beyond the local bay and, in severe cases, cascading failures or large-scale outages. This phase examines three representative attack strategies: (1) FDIA on SV from a Men-in-the-Middle (MitM) position, (2) FDIA via parallel SV injection without MitM access, and (3) FDIA targeting PTP synchronization. Each strategy exploits distinct vulnerabilities and challenges system resilience.

\begin{algorithm}[t]
\caption{MitM FDIA — force action}
\label{alg:force}
\small
\begin{algorithmic}[1]
\State Initialize interfaces: ${in\_if} \leftarrow eth0$, ${out\_if} \leftarrow eth1$
\State load params: scale, ramp, limits, log\_file
\While{running}
  \State pkt $\leftarrow$ capture(in\_if)
  \If{pkt.ethertype == 0x88BA and pkt.match(target)}
    \State sv $\leftarrow$ parse(pkt) \Comment{extract headers, samples}
    \State samples $\leftarrow$ modify(sv.samples, scale, ramp, limits)
    \State newpkt $\leftarrow$ build(pkt.headers, samples)
    \State wait\_to\_preserve\_timing(newpkt, sv.inter\_arrival)
    \State send(newpkt, out\_if); log(log\_file, pkt.meta)
  \Else
    \State forward(pkt, out\_if)
  \EndIf
\EndWhile
\end{algorithmic}
\end{algorithm}

In the MitM scenario, the adversary intercepts and manipulates SV traffic traversing the compromised link. Algorithm~\ref{alg:force} illustrates how SV streams may be altered to induce incorrect protection actions while preserving stealth. Traffic is filtered based on EtherType 0x88BA, and reconnaissance-derived parameters are used to modify SV payloads in a protocol-compliant manner. By emulating realistic abnormal conditions, such as short-circuit faults, the attacker can trigger relay misoperation without raising immediate alarms.

\begin{algorithm}[b]
\caption{FDIA from MitM — masquerade data}
\label{alg:masquerade}
\small
\begin{algorithmic}[1]
\State Initialize interfaces: ${in\_if} \leftarrow eth0$, ${out\_if} \leftarrow eth1$
\State load params: thresholds, limits, log\_file
\While{running}
    \State pkt $\leftarrow$ capture(in\_if)
    \If{pkt.ethertype == 0x88BA and pkt.match(target)}
    \State Capture and save as base\_values[idx]
    \EndIf
    \If{samples.curr $>$ Imax or samples.volt $<$ Vmin}
      \State repl $\leftarrow$ base\_values[idx]
      \State wait\_to\_preserve\_timing(newpkt, sv.inter\_arrival)
      \State send(newpkt, out\_if); log(log\_file, pkt.meta)
      \State idx $\leftarrow$ (idx + 1) mod thresholds
    \Else
    \State forward(pkt, out\_if)
    \EndIf
\EndWhile
\end{algorithmic}
\end{algorithm}

The same MitM position may also enable a masquerading attack aimed at concealing real faults, as shown in Algorithm~\ref{alg:masquerade}. In this case, the adversary records SV frames during normal operation to establish valid measurement profiles. When incoming frames exceed predefined thresholds, they are replaced with previously recorded “normal” data, masking genuine faults and preventing relay operation. As a result, faults may persist undetected, increasing the risk of equipment damage and subsequent cascading disconnections.

In a parallel injection scenario, the adversary injects falsified SV frames alongside legitimate traffic without requiring MitM access. As shown in Algorithm~\ref{alg:replay}, captured SV frames are stored and replayed, in some cases modifying only the sequence-of-data field containing measurement values. This creates ambiguity at the receiving IEDs, which may block protection functions or prioritize falsified frames over legitimate ones \cite{mocanu2021real}, achieving effects comparable to a MitM-based FDIA.

\begin{algorithm}[t]
\caption{Parallel FDIA — inject forged SV stream}
\label{alg:replay}
\small
\begin{algorithmic}[1]
\State Initialize interfaces: ${in\_if} \leftarrow eth0$, ${out\_if} \leftarrow eth0$
\State load params: replay\_rate, replay\_len, log\_file
\While{running}
  \State pkt $\leftarrow$ capture(in\_if)
  \If{pkt.ethertype == 0x88BA and pkt.match(target)}
    \State sv $\leftarrow$ parse(pkt) \Comment{headers, samples}
    \State store\_recent(sv)
  \EndIf
  \If{time\_to\_replay(replay\_rate)}
    \For{$k=1$ to replay\_len}
      \State newpkt $\leftarrow$ build(pkt)
      \State enforce\_timing(newpkt, rate\_limit)
      \State send(newpkt, out\_if)
      \State idx $\leftarrow$ (idx + 1) mod replay\_len
    \EndFor
  \EndIf
\EndWhile
\end{algorithmic}
\end{algorithm}

In the FDIA scenario targeting PTP, the adversary compromises the time synchronization mechanism that underpins coordinated protection behavior. Time synchronization is essential for correct event sequencing across IEDs. By launching a DoS attack against the GM, the attacker can delay or block timestamp distribution, causing IEDs to lose synchronization. In response, publishers of SV frames may reset the smpCnt field to indicate loss of synchronization, disrupting protection logic.

Substation networks may include multiple PTP paths from several GMs; however, in the considered scenario, the attacker’s objective is to compromise the legitimate GM rather than individual receivers \cite{alghamdi2021precision}. To increase the impact, the adversary may inject counterfeit PTP packets broadcasting incorrect time values in parallel with legitimate messages, as illustrated in Algorithm~\ref{alg:ptp}. The resulting desynchronization among PTP-dependent devices can block protection functions and leave the substation exposed to real faults, with consequences comparable to previous scenarios.

\begin{algorithm}[t]
\caption{PTP FDIA — spoof / flood}
\label{alg:ptp}
\small
\begin{algorithmic}[1]
\State Initialize interfaces: ${in\_if} \leftarrow eth0$, ${out\_if} \leftarrow eth0$
\State load params: inject\_rate, inject\_len, log\_file
\While{running}
  \State pkt $\leftarrow$ capture(in\_if)
  \If{pkt.ethertype == 0x88F7 and pkt.match(target)}
    \State ptp $\leftarrow$ parse\_PTP(pkt)
    \State store\_recent(ptp)
  \EndIf
  \If{time\_to\_inject(inject\_rate)}
    \For{$k=1$ to inject\_len}
        \State newpkt $\leftarrow$ build\_PTP(repl, local\_time\_offset)
        \State newpkt $\leftarrow$ build\_PTP(flood\_template)
      \State send(newpkt, out\_if); log(log\_file)
      \State idx $\leftarrow$ (idx + 1) mod inject\_len
    \EndFor
  \EndIf
\EndWhile
\end{algorithmic}
\end{algorithm}

\subsection{Phase 4 — Impact of cyberattacks}

The final phase addresses the impact of the attack and the adversary’s efforts to remain undetected. Consequences may range from localized service disruptions to large-scale grid instability. Two representative outcomes are considered: maintaining stealth to prolong attacker presence and assessing the resulting operational damage. Together, they illustrate the challenges associated with detecting and mitigating advanced cyber threats.

In the first outcome, the adversary prioritizes evasion to extend access to the substation environment. This may include relocating malicious hardware or software to alternative nodes, replicating attack logic across compromised devices, or remotely erasing traces from engineering workstations. Such actions obscure indicators of compromise—such as unauthorized configuration changes or falsified measurements—and complicate forensic analysis, enabling continued exploitation without triggering immediate defensive responses.

In the second outcome, the adversary evaluates system behavior following the attack. This may manifest as either no action or incorrect action by protection devices. In the no-action case, genuine faults are masked by falsified “normal” measurements, allowing abnormal conditions to persist and cause equipment damage. In the wrong-action case, fabricated fault data initiates unnecessary trips, disconnecting healthy components and destabilizing power flows. Manipulated MMS data reaching higher-level systems may further propagate erroneous control actions, amplifying disturbances and increasing the risk of cascading outages.

Beyond immediate operational effects, the adversary may refine future attack strategies based on observed system responses or exploit reconnaissance data—such as device configurations or measurement profiles—for secondary objectives. 

Evaluating the feasibility and consequences of such attacks is most effectively conducted in a controlled laboratory environment using industrial-grade equipment. Experimental validation is therefore essential for understanding the impact of advanced FDIAs, assessing countermeasures, and strengthening the resilience of IEC~61850-based substation automation systems.

\section{Results and Analysis}\label{sec:results_analysis}
\subsection{Cyber-Physical Testbed Architecture}

To evaluate advanced FDIAs against the SV protocol, an isolated cyber-physical testbed was developed to reproduce bay-level operation using industrial-grade equipment. The testbed provides controlled and repeatable conditions for attack execution and measurement \cite{mishch}. The evaluation focuses exclusively on the bay level—the lowest digital layer of a substation—therefore avoiding the need to model a wider grid.

\begin{figure}[!htbp]
\centering
\includegraphics[width=1\columnwidth]{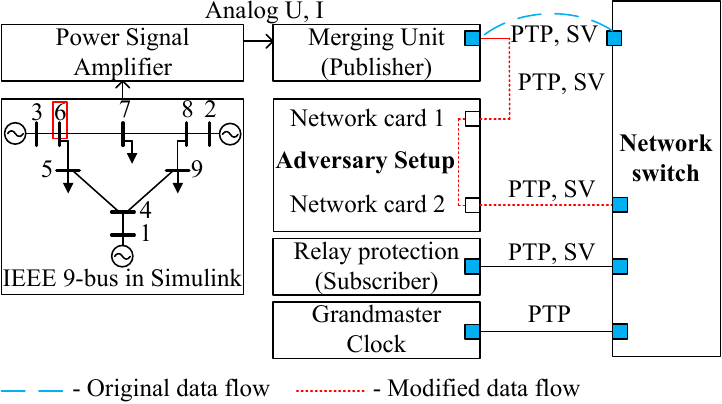}
\caption{Cyber-physical testbed for analyzing advanced FDIAs at the bay level}
\label{fig:Setup}
\end{figure}

The IEEE~9-bus system is adopted as the power system reference, with experiments centered on bus~6. Fig.~\ref{fig:Setup} illustrates the hardware configuration and signal flow for this bus. A Simulink model is used for offline generation of current and voltage waveforms for the line between buses~6 and~7; it is not part of the real-time protection loop. These waveforms are applied to an OMICRON CMC~356 power signal amplifier, which generates physical analog signals.

The amplifier feeds analog measurements to a Siemens SIPROTEC~5 (7SX85) IED configured to operate as a MU and publish SV frames at 4000~Hz. A second SIPROTEC~5 protection relay from the same vendor subscribes to the SV stream and implements distance protection for the line between buses~6 and~7. The relay trip output is hardwired back to the amplifier input, enabling closed-loop testing with real protection logic operating in real time.

Time synchronization is provided by a GM implemented on a dedicated workstation running \texttt{ptp4l} with hardware timestamping enabled \cite{linuxptp2023}. The GM ensures that the global synchronization flag is set in the smpCnt field of SV frames, which is required for correct protection operation.

The adversary setup is implemented on a workstation equipped with two network interface cards and a low-latency C++ framework for packet interception and modification. One interface connects to the MU and the other to the process bus switch, enabling MitM redirection of SV traffic through the adversary setup. The workstation operates on Kali Linux, running penetration-testing tools alongside custom-developed code optimized to satisfy SV timing constraints. \cite{kalilinux2025}.

Following initial access, the adversary performs passive and active reconnaissance to extract node identifiers, sampling rates, smpCnt behavior, frame formats, CT/VT ratios, and protection settings \cite{PowerTech2025}. These observations are used to construct protocol-compliant SV frames that maximize stealth.

\subsection{Scenario 1 — FDIA from MitM (force the action)}

This experiment evaluates whether an adversary positioned as a MitM at the bay level can inject falsified SV data that emulates a close-in three-phase short-circuit and provokes an undesired relay trip. The attacker is assumed to rely solely on bay-level knowledge obtained through prior reconnaissance.

Falsified SV samples were constructed using reconnaissance-derived observations and a limited set of realistic assumptions. Captured SV traffic indicated a normal secondary peak current of approximately 0.370~A and, from configuration data, a CT ratio of 600/1. Modeling a close-in low-impedance fault as a 0.1~$\Omega$ condition—a typical value for such events—yields an estimated secondary peak current of approximately 22~A \cite{kasikci2018short}. During the simulated fault, voltage was reduced to approximately 5~V, consistent with close-in short-circuit behavior \cite{anderson2021power}. These values represent conservative and plausible fault magnitudes given the attacker’s limited system visibility.

\begin{figure}[t]
\centering
\includegraphics[width=1\columnwidth]{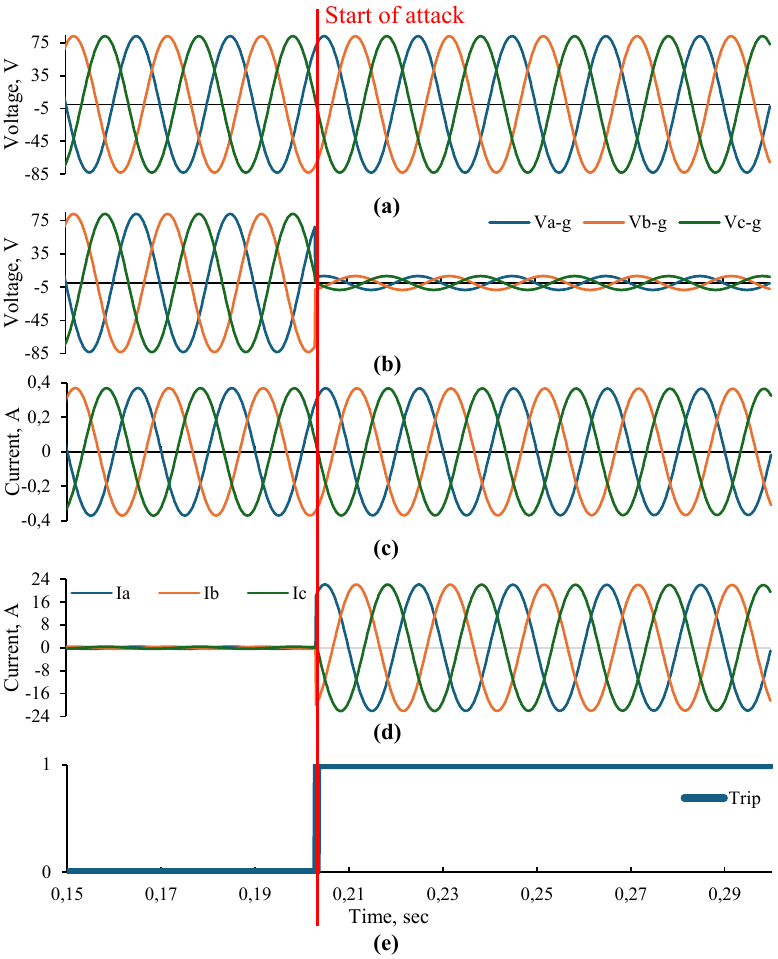}
\caption{Creation of a false short circuit. (a) Original voltage waveform published by the MU; (b) Voltage waveform received by the relay; (c) Original current waveform published by the MU; (d) Current waveform received by the relay; (e) Relay response showing trip signal}
\label{fig:fake}
\end{figure}

Developed C++ application running on a dual-NIC workstation intercepted SV frames on one interface, parsed the current and voltage samples, replaced them with the synthesized fault values, preserved all SV headers and timing characteristics, and forwarded the modified frames through the second interface.

\begin{figure}[t]
\centering
\includegraphics[width=1\columnwidth]{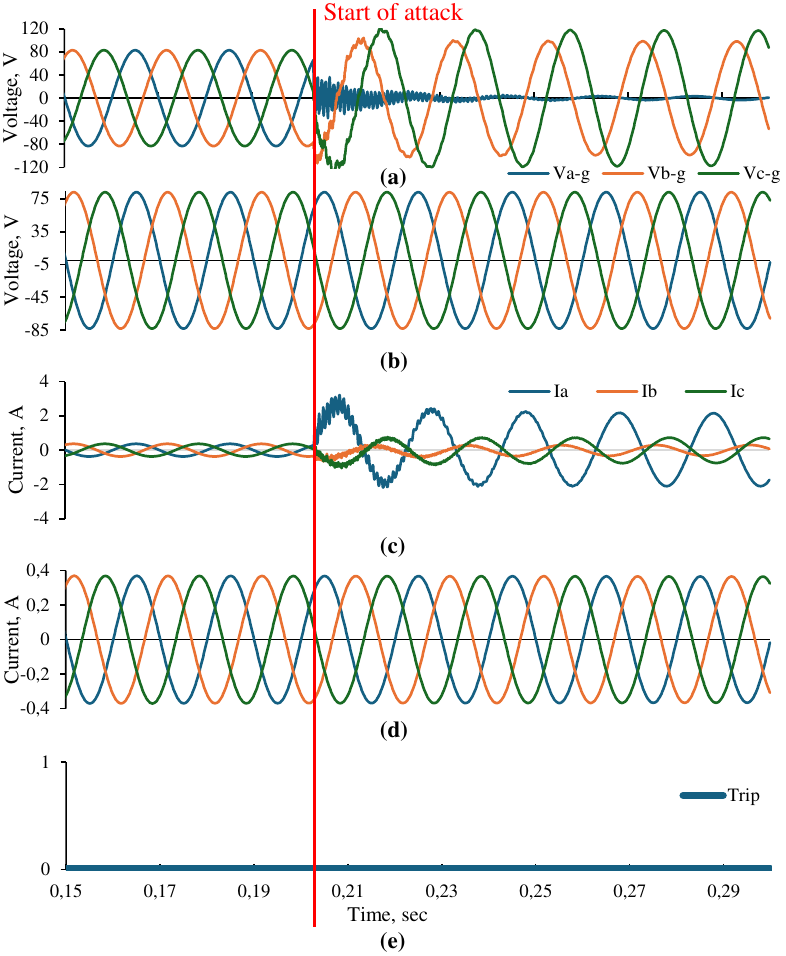}
\caption{Masquerading a short-circuit fault: (a) Original voltage waveform published by the MU; (b) Voltage waveform received by the relay; (c) Original current waveform published by the MU; (d) Current waveform received by the relay; (e) Relay response showing absence of trip signal}
\label{fig:Real}
\end{figure}

The injected SV stream successfully triggered the relay’s distance protection, resulting in an undesired trip without raising alarms or integrity warnings. The trip was recorded by both the protection relay and the power signal amplifier. Fig.~\ref{fig:fake} shows the injected waveform together with the relay response. This outcome confirms that a carefully crafted MitM FDIA can induce stealthy protection misoperation under realistic bay-level assumptions.

\subsection{Scenario 2 — FDIA from MitM (masquerade data)}

This experiment investigates whether an adversary positioned as a MitM at the bay level can conceal a real short-circuit by replacing fault-related SV samples with values reflecting normal system operation, thereby preventing fault detection and isolation. The same MitM position and hardware configuration as in Scenario~1 were used; only the attack logic differed, relying on recorded benign samples rather than synthesized fault values.

During an initial learning phase, the adversary captured SV frames under normal dynamic conditions and derived baseline thresholds for current and voltage. For the experiments reported here, the observed secondary peak current was approximately 0.370~A and the representative phase-to-ground voltage was 83~V, both well below fault levels. The C++ implementation continuously parsed incoming SV frames, compared measured values against the stored thresholds, and, when these bounds were exceeded during a fault, replaced the live samples with previously recorded baseline frames.

A single-phase-to-ground fault was introduced in Simulink and reproduced through the power signal amplifier \cite{anderson2021power, kasikci2018short}. At fault inception, the masquerading logic detected abnormal values and substituted baseline SV frames before forwarding them to the relay. As a result, the relay continued to observe nominal measurements and did not operate; no trip was issued and no binary signals were registered at the amplifier. Fig.~\ref{fig:Real} presents the masked waveforms together with the relay’s non-response.

This scenario demonstrates that a MitM-based masquerading FDIA can successfully suppress bay-level protection operation, allowing real faults to persist undetected and increasing the risk of equipment damage and cascading outages.

\subsection{Scenario 3 — Blocking protection functions by replaying data}

This experiment evaluates whether an adversary with only unauthorized access to the process bus can disable protection functions by injecting a parallel SV stream, without requiring inline MitM control. The attacker is assumed to have bay-level access through a switch port and to operate within the same isolated environment used in the preceding scenarios.

In this setup, the adversary’s workstation was connected to an unused switch port while legitimate SV traffic continued along its original path. SV frames were passively captured and used to construct a forged SV stream. To remain stealthy, the injected frames preserved all protocol headers and timing characteristics and differed only in the measurement payloads.

Developed C++ application stored recently observed SV samples and replayed them as a parallel multicast stream. During injection, forged frames replicated the original publisher’s header attributes, causing subscribers to receive two concurrent SV streams with identical identifiers but inconsistent measurement values.

\begin{figure}[!tb]
\centering
\includegraphics[width=1\columnwidth]{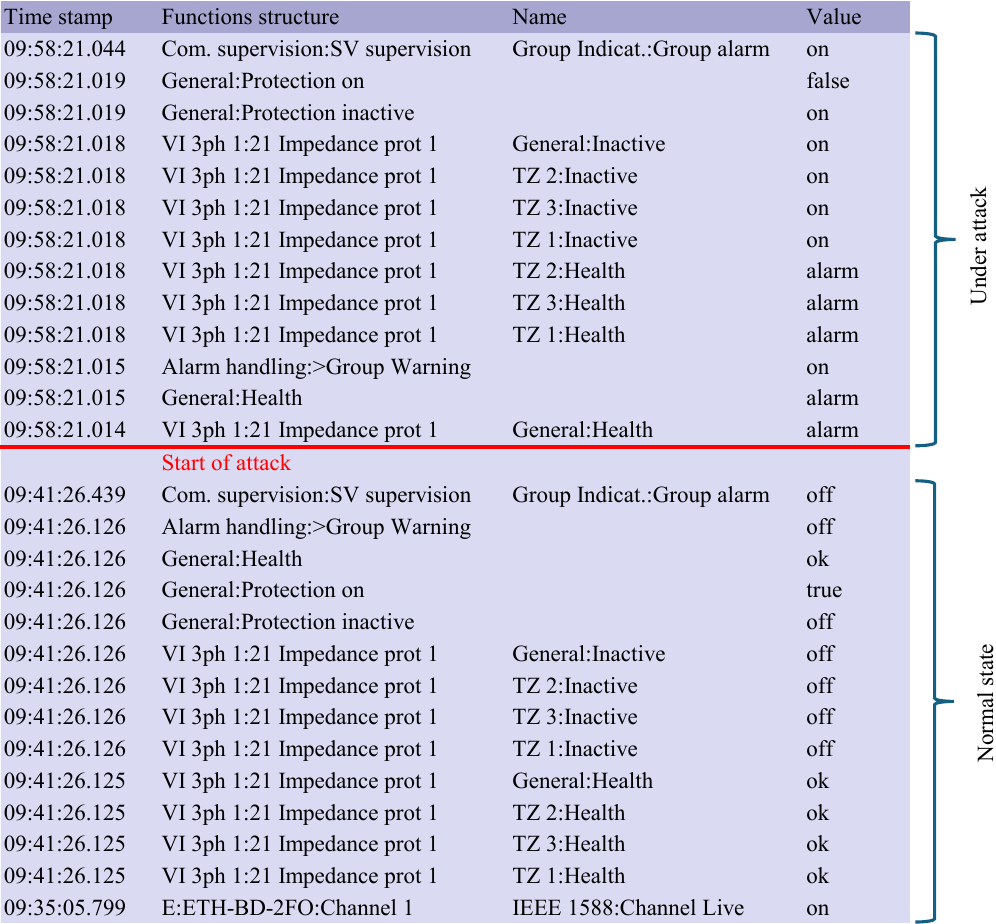}
\caption{Relay event log during a replay attack, showing blocking of protection functions after detection of conflicting SV streams}
\label{fig:parallel}
\end{figure}

Upon detecting this inconsistency, the protection relay raised internal alarms and entered a blocking state to prevent operation on unreliable inputs. This behavior was logged by the relay and resulted in suppression of protection functions, such that no trip would occur even in the presence of a genuine fault, as shown in Fig.~\ref{fig:parallel}.

These results demonstrate that parallel SV injection can force IEDs into fail-safe blocking, leaving real faults unmitigated and increasing the risk of equipment damage and fault propagation. The scenario underscores the need for bay-level protections capable of detecting and mitigating parallel-stream manipulation on the process bus.

\subsection{Scenario 4 — FDIA targeting time synchronization}

The objective of this experiment was to assess whether an adversary could disable protection functions by corrupting PTP synchronization rather than manipulating SV traffic. Accurate time reference is essential for coordinated measurements and correct operation of protection logic across MUs and relays.

The testbed configuration was identical to Scenario~3, with the adversary connected to the process bus through a single switch port. Instead of altering SV traffic, the attack targeted time synchronization using two complementary methods. First, the GM configuration was modified to publish PTP messages at a rate insufficient for IED requirements, emulating delay effects consistent with a DoS condition \cite{akbarzadeh2023attacking}. Second, the adversary injected a parallel PTP stream using \texttt{ptp4l}, introducing conflicting time sources on the bus.

Both the MU and the protection relay detected synchronization inconsistencies. The relay logged synchronization errors, while the MU marked outgoing SV frames with a “none” synchronization status instead of “global,” indicating loss of valid reference time. In response, the relay automatically blocked protection functions. When a real fault occurred under these conditions, no trip was issued and the fault persisted, replicating the operational impact observed in Scenario~3 and increasing the risk of equipment damage and fault propagation.

This experiment demonstrates that compromising auxiliary services such as PTP can disrupt protection operation as effectively as direct SV manipulation. Even when measurement traffic remains intact, loss of time synchronization undermines the reliability of IEC~61850-based protection, highlighting the importance of securing both primary data flows and supporting services.

Across all four scenarios, the experiments show that advanced FDIAs at the bay level can compromise both measurement traffic and auxiliary services. By injecting falsified SV frames, replaying valid data, or corrupting time synchronization, an adversary can provoke false trips, mask real faults, or block protection functions. These outcomes risk equipment damage, fault propagation, and cascading failures. Importantly, most attacks operated without generating alarms or logs, revealing the limitations of existing defenses. The findings underscore the need for protection strategies that combine resilience against SV manipulation with safeguards for synchronization and anomaly detection.

\section{Proposed Measures}
\label{sec:proposed_measures}

The coordinated FDIAs demonstrated in Scenarios~1–4 highlight the need for protection strategies that remain effective even when primary communication channels are compromised. Such measures must satisfy strict real-time constraints, operate autonomously at the bay level, and preserve protection functionality during active attacks. This section first analyzes existing defense mechanisms within a defense lifecycle perspective and identifies their limitations in the context of coordinated and physically consistent attacks. Based on this analysis, a resilience-oriented approach is introduced to maintain correct protection operation under compromised conditions.

\subsection{Defense Lifecycle Analysis}

\begin{figure}[!b]
\centering
\includegraphics[width=1\columnwidth]{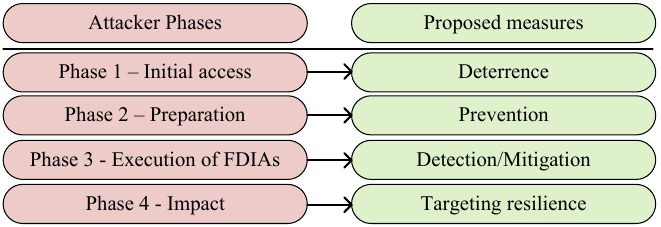}
\caption{Defense strategy mapped to the four phases of advanced FDIA scenarios}
\label{fig:measures}
\end{figure}

Advanced cyber-physical attacks cannot be mitigated by a single mechanism. A defense approach spanning the full attack lifecycle is required. Figure~\ref{fig:measures} maps representative defense measures to the four phases of coordinated FDIAs considered in this work, following the lifecycle model in~\cite{ashok2017cyber}, highlighting their practical limitations in the context of coordinated FDIAs.

\textbf{Deterrence, Phase 1 -- Initial Access:} 
Physical access control, surveillance, tamper detection, and network segmentation reduce the likelihood of initial compromise and reconnaissance~\cite{mahato2024physical,hussain2023effective, orr2021securing}. However, these measures are less effective against insider threats or attackers with legitimate access, such as compromised engineering workstations.

\textbf{Prevention, Phase 2 -- Preparation:} 
Security practices such as patch management, penetration testing, and encryption aim to eliminate vulnerabilities before attack execution~\cite{searle2016nescor}. Advanced techniques such as moving target defense further limit attacker effectiveness by dynamically altering communication parameters, thereby reducing predictability and exposure~\cite{zhang2021smart}. Nevertheless, strict real-time constraints and the multicast nature of SV communication limit the applicability of computationally intensive mechanisms such as full encryption and frequent updates at the process bus.

\textbf{Detection and Mitigation, Phase 3 -- Execution:} 
Detection mechanisms, including intrusion detection systems and state estimation techniques, aim to identify anomalies and trigger mitigation actions~\cite{musleh2019survey}. In practice, centralized methods are often unsuitable for bay-level deployment, while lightweight approaches may fail to detect coordinated attacks that preserve normal traffic characteristics and physical consistency.

\textbf{Targeting Resilience, Phase 4 -- Impact:} 
Resilience mechanisms aim to maintain correct system operation even when attacks are not detected or prevented. In the context of SV-based protection, this requires independent validation of measurement data and protection decisions, ensuring that compromised communication channels do not directly lead to incorrect relay operation.

Taken together, these measures form a complementary defense framework spanning the attack lifecycle. However, their effectiveness depends on deployment constraints and the characteristics of the attack. In particular, coordinated and physically consistent FDIAs challenge conventional assumptions used in prevention and detection mechanisms. This motivates the need for resilience-oriented approaches that ensure correct protection operation even when primary communication channels are compromised. 

\subsection{Resilience-Oriented Protection Using Independent Channels}

\begin{figure}[!b]
\centering
\includegraphics[width=1\columnwidth]{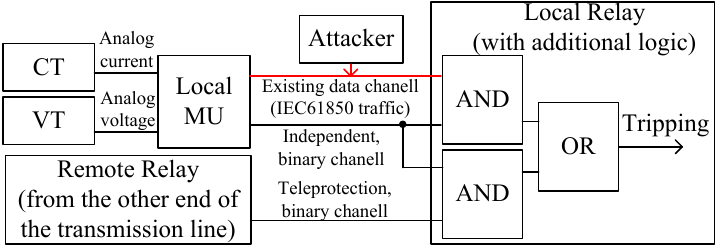}
\caption{Enhanced protection logic with MU-derived independent channel and teleprotection relay confirmation to counter FDIAs at the process bus}
\label{fig:Resilience}
\end{figure}

The proposed approach focuses on resilience rather than prevention alone, recognizing that advanced attackers may bypass or disable conventional safeguards. The objective is to sustain essential protection functions during an attack, not merely to restore operation afterward. Figure~\ref{fig:Resilience} illustrates the protection logic extension designed to counter all analyzed scenarios. The method introduces cross-verification of SV data using:
\begin{itemize}
    \item an independent, trusted binary channel between the MU and the protection relay, and
    \item an independent, trusted device—such as a relay at the opposite end of the line—connected to the protection relay.
\end{itemize}

The analog signals measured by CTs and VTs cannot be altered through cyber means prior to digitization. The MU, which converts these signals into digital samples, therefore constitutes the first trustworthy element in the measurement chain. A binary signal derived directly from the MU can be transmitted via a physically and logically isolated channel, independent of IEC~61850 communication. The protection relay may then issue a trip only when two conditions are simultaneously satisfied: (1) reception of valid SV frames over the process bus, and (2) confirmation via the MU’s independent binary channel. This dual validation significantly constrains the feasibility of Scenario~1, as an attacker would need to compromise both channels concurrently.

For Scenarios~2–4, where one communication path may be disrupted or rendered unreliable, the MU channel alone is insufficient. To address this, the protection logic can be extended to incorporate confirmation from an independent, trusted relay located at the opposite end of the protected line. Using established pilot protection (teleprotection) mechanisms~\cite{hou2008iec}, the two relays exchange status signals over a separate path. With minor adaptation, teleprotection can serve as an additional resilience channel, allowing the relay under attack to trip even when local SV measurements are unavailable or inconsistent, provided that: (1) a valid teleprotection signal is received from the remote relay, and (2) confirmation is obtained from the local MU’s independent binary channel.

This resilience-oriented logic directly addresses all four attack scenarios. In Scenario~1, falsified SV data fail to produce a trip due to missing MU confirmation. In Scenarios~2–4, where SV or PTP manipulation disables local protection, the combined use of MU confirmation and remote relay signaling enables continued operation. Importantly, this approach does not replace the primary communication infrastructure but augments it, allowing improved resilience using existing devices and communication paths with minimal architectural modification.

From a deployment perspective, the proposed resilience extension requires minimal additional infrastructure, as it relies on hardware capabilities already available in modern substations. The MU-derived binary signal can be implemented using existing hardwired I/O, while remote-end confirmation can leverage standard teleprotection interfaces commonly installed on line and cable bays. As a result, the implementation cost is low and does not require modifications to the process-bus communication network. The approach does, however, require an extension of the protection logic within the relay firmware to incorporate multi-source verification and decision fusion, which can be realized as an incremental logic update rather than a redesign of core protection functions. The approach is inherently scalable, since each protected bay already includes an MU and a protection relay capable of providing independent confirmation channels. Furthermore, the latency introduced by hardwired binary signaling is typically in the sub-millisecond range, and teleprotection channels operate within the millisecond range, which is compatible with the timing requirements of distance and differential protection schemes. These properties indicate that the proposed resilience mechanism is practical, low-overhead, and suitable for deployment across a wide range of substation architectures.

\section{Conclusion and Future Work}\label{sec:conclusion}

Modern substations increasingly rely on real-time multicast communication for protection and automation, exposing new attack surfaces at the bay level where IEDs operate with limited supervision.

This study presented a scenario-based analysis of advanced FDIAs targeting IEC~61850 SV and PTP synchronization. In contrast to simplified attack representations, the considered scenarios implement coordinated multi-vector manipulation strategies that preserve physical consistency of electrical parameters. All attacks were validated on PHIL testbed using industrial-grade IEDs and a realistic communication network. The results demonstrate that such attacks can remain stealthy under real device constraints, bypass standard safeguards, and trigger false trips, conceal faults, or block protection actions without raising alarms.

These findings highlight that commonly assumed protection and detection mechanisms may not be sufficient when faced with coordinated and physically consistent attack behavior implemented under realistic substation conditions. In response, countermeasures were analyzed within a defense perspective, with particular emphasis on resilience at the bay level. A resilience-oriented protection extension based on independent trusted channels and cross-verification of protection decisions was proposed. By leveraging existing teleprotection principles, this approach enhances robustness against SV-based FDIAs without requiring fundamental changes to the primary communication infrastructure.

Future work will extend the analysis to coordinated multi-substation scenarios, including interactions with station-level and control-center systems. In addition, the proposed resilience-oriented mechanism will be further developed and experimentally validated with respect to real-time constraints, interoperability with existing protection schemes, and scalability across heterogeneous substations. Further research will also investigate lightweight detection mechanisms and complementary redundancy strategies to strengthen practical deployment.

\bibliographystyle{elsarticle-num} 
\bibliography{refs.bib}






\end{document}